\documentclass[conference]{IEEEtran}
\IEEEoverridecommandlockouts
\usepackage{cite}
\usepackage[T1]{fontenc}

\usepackage{amsmath,amssymb,amsfonts}
\usepackage{algorithm}
\usepackage{algpseudocode}
\usepackage{textcomp}
\usepackage{float}
\usepackage{soul}
\usepackage{xcolor}
\usepackage{multirow}
\usepackage{ulem}
\usepackage{booktabs}
\usepackage{graphicx}
\usepackage{threeparttable}
\def\BibTeX{{\rm B\kern-.05em{\sc i\kern-.025em b}\kern-.08em
    T\kern-.1667em\lower.7ex\hbox{E}\kern-.125emX}}

\begin{document}

\title{Poison in the Well: Feature Embedding Disruption in Backdoor Attacks\\

}


\author{
Zhou Feng\textsuperscript{1$\dagger$}, Jiahao Chen\textsuperscript{1$\dagger$}, Chunyi Zhou\textsuperscript{1*}, Yuwen Pu\textsuperscript{2}, Qingming Li\textsuperscript{1}, Shouling Ji\textsuperscript{1} \\
\textsuperscript{1}College of Computer Science and Technology, Zhejiang University, Hangzhou, China \\
\textsuperscript{2}School of Big Data \& Software Engineering, Chongqing University, Chongqing, China \\
\{zhou.feng, xaddwell, zhouchunyi, liqm, sji\}@zju.edu.cn, yw.pu@cqu.edu.cn
}


\maketitle

\begingroup
\renewcommand\thefootnote{}
\footnotetext{
\hspace*{-\parindent}\textsuperscript{$\dagger$}Zhou Feng and Jiahao Chen contributed equally to this work. \par
\hspace*{-\parindent}\textsuperscript{*}Corresponding author: Dr. Chunyi Zhou (zhouchunyi@zju.edu.cn)
}
\endgroup

\begin{abstract}

Backdoor attacks embed malicious triggers into training data, enabling attackers to manipulate neural network behavior during inference while maintaining high accuracy on benign inputs. 
However, existing backdoor attacks face limitations manifesting in excessive reliance on training data, poor stealth, and instability, which hinder their effectiveness in real-world applications.
Therefore, this paper introduces \textit{ShadowPrint}, a versatile backdoor attack that targets feature embeddings within neural networks to achieve high ASRs and stealthiness. Unlike traditional approaches, \textit{ShadowPrint} reduces reliance on training data access and operates effectively with exceedingly low poison rates (as low as 0.01\%). It leverages a clustering-based optimization strategy to align feature embeddings, ensuring robust performance across diverse scenarios while maintaining stability and stealth. 
Extensive evaluations demonstrate that \textit{ShadowPrint} achieves superior ASR (up to 100\%), steady CA (with decay no more than 1\% in most cases), and low DDR (averaging below 5\%) across both clean-label and dirty-label settings, and with poison rates ranging from as low as 0.01\% to 0.05\%, setting a new standard for backdoor attack capabilities and emphasizing the need for advanced defense strategies focused on feature space manipulations.
\end{abstract}


\section{Introduction}

Rapid advancement of deep learning has led to its adoption in various domains, from autonomous vehicles and healthcare to finance and security systems~\cite{shen2017deep,heaton2017deep}. However, the increasing reliance on neural networks has also exposed them to a variety of adversarial threats. Among these, backdoor attacks~\cite{gu2019badnets} have emerged as a particularly stealthy and potent form of vulnerability. By embedding a malicious trigger into the training data, an attacker can cause the model to produce incorrect or harmful outputs in the presence of the trigger, while maintaining high accuracy on benign inputs.

The effectiveness of backdoor attacks often hinges on the design and placement of the trigger, as well as the method's ability to evade detection. 
Recent advances in backdoor attacks have introduced innovative approaches that address some of these challenges. For example, Narcissus~\cite{zeng2023narcissus}, require only access to training data from the target class, which makes them less dependent on extensive data access. Or like Gao et al.~\cite{gao2023not}, identify and exploit ``hard" samples, data with weak robust features, to complement existing clean-label attacks. Although these methods represent significant progress, they still share several inherent limitations that constrain their applicability and effectiveness:

\begin{itemize}
    \item \textbf{Over-Reliance on Data Access:} Many existing approaches often assume attackers to have extensive or unrestrained knowledge of training data and demand excessively high poison rates.
    In real-world scenarios, this access is often partial or severely limited, making these assumptions impractical.
    \item \textbf{Inadequate Stealthiness:} The invisibility of the trigger remains a critical factor, as detectable triggers are highly susceptible to identification
    by defense mechanisms.
    \item \textbf{Fragile Stability:} Achieving consistent performance across various input scenarios (e.g., white-box, black-box and data-free, as described in Section \ref{Threat Model}) and ensuring resilience to variations in data distribution continue to be significant challenges.
\end{itemize}

The SOTA methods face additional challenges when applied to MLaaS platforms~\cite{kim2018nsml, 2015mlaas}. These platforms, widely used in real-world applications, present a unique set of constraints, including limited access to the model architecture, training data, and hyperparameters. Many existing backdoor attacks falter in these environments due to their dependence on high levels of attacker knowledge and access. This underscores the importance of designing attacks that are adaptable to real-world scenarios, where constraints and defenses are more robust than in traditional experimental settings.

Therefore, this paper aims to address the above gaps by investigating novel ways to analyze and exploit the relationships within the feature space during backdoor attacks. Specifically, the contributions of this work are as follows:

\begin{itemize}
    \item We propose \textit{ShadowPrint}, a novel backdoor attack that 
    mitigates assumptions about attacker and achieves robust attack performance under realistic countermeasures.

    \item We employs a clustering-based trigger optimization strategy to align feature embeddings of poisoned samples, reducing the burden of backdoor learning during model training, enabling the use of an extremely low poison rate (as low as $0.01\%$) while maintaining attack effectiveness.

    \item 
    
    We conduct extensive experiments on multiple benchmark datasets (i.e., CIFAR-10, CIFAR-100, and TinyImageNet),
    demonstrating that \textit{ShadowPrint} achieves high attack performance. Specifically, it maintains effectiveness, stealthiness, and stability while evading detection under SOTA defenses, such as IBD-PSC~\cite{hou2024ibd}, SCALE UP~\cite{guo2023scale}, and Beatrix~\cite{ma2022beatrix}, even at extremely low poison rates (no greater than $0.05\%$).

\end{itemize}

\section{Background and Related Work}

\subsection{Backdoor Attacks in Deep Learning}

Backdoor attacks pose a unique and stealthy threat to neural networks~\cite{gu2019badnets, chen2017targeted, nguyen2021wanet, liu2018trojaning, turner2019label, nguyen2020input, barni2019new}. 
Early techniques, like BadNets~\cite{gu2019badnets}, introduced the concept of injecting simple and static triggers into training samples to induce misclassification.
Subsequent approaches~\cite{chen2017targeted, nguyen2021wanet, liu2018trojaning, turner2019label, nguyen2020input, barni2019new}, including Blended Attacks~\cite{chen2017targeted}, sought to enhancing the stealthiness by blending imperceptible patterns into input data. 
However, such methods predominantly focus on the superficial properties of triggers, such as their appearance, and neglect a deeper investigation of how the triggers interact with the model's feature space, leaving room for further refinement. 

Recently, researchers have explored data-agnostic techniques and adaptive trigger designs~\cite{zeng2023narcissus, li2023efficient, gao2023not, wu2023computation, zhu2023boosting}, aiming to generalize across datasets and architectures. 
Li et al.~\cite{li2023efficient} propose an efficient data-constrained backdoor attacks, reflecting practical conditions where attackers only have partial access to training data. Similarly, computationally informed strategies such as We et al.~\cite{wu2023computation} propose novel metrics to select poisoned samples that are more effective in reshaping decision boundaries, while Zhu et al.~\cite{zhu2023boosting} present a learnable strategy to poison sample selection using a min-max optimization framework. 
Though these methods enhance flexibility and stealthiness, they often prioritize the trigger’s superficial properties over a deeper exploration of the model's internal feature representations. Key limitations, including over-reliance on data access, inadequate stealth for triggers, and lack of stability, underscore the need for innovative strategies that address these gaps. 
\subsection{Backdoor Defensive Mechanisms}

Defensive mechanisms against backdoor attacks can be grouped into three categories: data sanitation~\cite{wang2019neural, gao2019strip}, model inspection~\cite{chen2018detecting, liu2018fine, ma2022beatrix}, and runtime detection~\cite{hou2024ibd, guo2023scale}. 

Data sanitation methods, such as Neural Cleanse~\cite{wang2019neural} and STRIP~\cite{gao2019strip}, attempt to identify and remove poisoned samples from the training data. Although these methods can be effective against certain trigger types, they struggle with detecting more adaptive or imperceptible triggers that evade traditional detection strategies.
Model inspection techniques focus on analyzing the model’s internal behavior to identify anomalies that may signal the presence of a backdoor. Activation clustering~\cite{chen2018detecting} and gradient-based analysis~\cite{liu2018fine} are examples of such approaches that seek to differentiate between clean and poisoned samples. More recently, Beatrix~\cite{ma2022beatrix} introduced a novel technique for identifying poisoned samples by analyzing activation anomalies via Gram matrices. However, these methods face scalability and generalizability challenges, particularly when applied to more complex models.
Runtime detection defenses flag suspicious inputs during training process. Approaches like IBD-PSC~\cite{hou2024ibd} and SCALE UP~\cite{guo2023scale}, aim to detect the presence of triggers at the input level.

Despite advancements in these defense mechanisms, the evolving sophistication of backdoor attacks continues to outpace current methodologies. 


\section{Methodology}


\subsection{Threat Model} \label{Threat Model}

To systematically evaluate \textit{ShadowPrint}, we categorize attackers into three types based on their capabilities, ensuring coverage of diverse real-world attack scenarios. These attacker types are meaningful as they reflect varying levels of knowledge and resources available to adversaries:

\begin{itemize}
    \item \textbf{Scenario A1 (White-Box):} The attacker has knowledge of the model architecture and partial training dataset.
    \item \textbf{Scenario A2 (Black-Box):} The attacker only has knowledge of partial training dataset.
    \item \textbf{Scenario A3 (Data-Free):} The attacker has no knowledge of the model architecture or the training dataset.
\end{itemize}

Note that all these attackers can manipulate an extremely small number of training samples for poisoning (e.g., less than $0.05\%$ in this paper, which means that less than 25 samples manageable in training dataset like CIFAR10 with 50000 training samples in total, and they can launch both clean label and dirty label attacks considering their specific capability. Their common objective is to maintain the model's accuracy on clean samples while ensuring high success rates on poisoned samples.


\subsection{Overview of ShadowPrint}


\begin{figure}[h]
    \centering
    \includegraphics[width=0.8\linewidth]{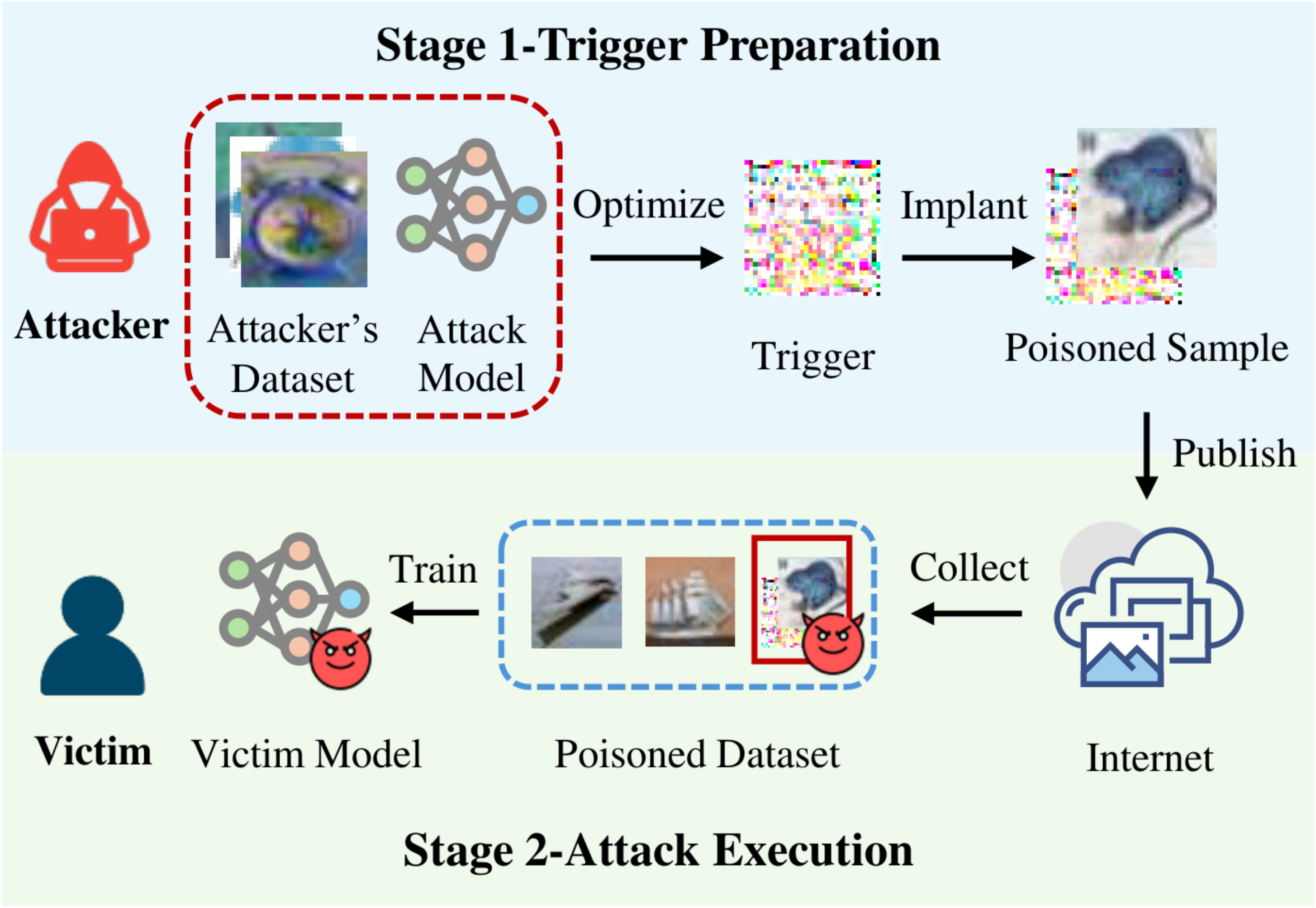}
    \caption{Scheme of \textit{ShadowPrint} framework. In Stage 1, the attacker optimizes the trigger based on his knowledge and resources. In Stage 2, the victim, unaware of the malicious intent, uses the poisoned dataset provided via the Internet to unintentionally train a backdoored model.}
    \label{fig:scheme}
\end{figure}

\textit{ShadowPrint} introduces a novel backdoor attack framework that leverages feature space manipulation for enhanced stealth and effectiveness. By optimizing a backdoor trigger to align poisoned samples in the embedding space, it reduces reliance on extensive training data access and high poison rates. Unlike traditional methods that focus on visible or statistical anomalies, \textit{ShadowPrint} directly targets the model’s internal representations to ensure stability and resilience across diverse scenarios. The overall scheme is illustrated in Fig.~\ref{fig:scheme}.

\subsection{Method Design}
\subsubsection{Stage 1-Trigger Preparation}

Let $f$ represent the target neural network with parameter $\theta $, trained on a dataset $\mathcal{D} = \{(x_i, y_i)\}_{i=1}^N $, where $x_i $ and $y_i $ denote the input samples and their corresponding labels, respectively. The goal of \textit{ShadowPrint} is to inject a backdoor with trigger $t$ into the parameters $\theta$  by utilizing the attacker's knowledge of the model and dataset, such that poisoned samples $x'_i = T(x_i, t) $ infiltrate the dataset $\mathcal{D} $ and disrupt the decision-making process of $f_\theta $, where $T$ is the transformation function. Specifically:
\begin{itemize}
    \item The model misclassifies $x'_i $ to a target label $y_t $.
    \item The accuracy on clean samples remains unaffected.
    \item Evade the potential detection by defenders.
\end{itemize}

Here in this paper, for attack stealthiness, we adopt the transformation function $T$ in line with the Blended attack~\cite{chen2017targeted}:
\begin{equation}
    T(x_i, t) = x_i\times (1-w) + t\times w
\label{eq:blended}
\end{equation}
where $w$ denotes the trigger weight and the size of trigger $t$ is the same as the samples.

With the goal and formulation above, we start with the main goal of a backdoor attack, which means minimizing $\mathcal{L}(f(T(x_i,t)), y_t)$. However, we propose that this backdoor behavior can also be expressed as:
\begin{equation}
    \min \sum_{i,j \, :\, i \neq j} D(f(T(x_j,t)), f(T(x_i,t))) + \mathcal{L}(f(T(x_i,t)),y_t)
\end{equation}
where the first term stands for backdoor clustering that aims to minimize the distance (use measurement $D$) of the triggered samples $T(x_i,t)$ and the second term specifies the target class $y_t$ for backdoor attack. With this analysis, we can reformulate the learning of the backdoor attack as a clustering optimization, and the cluster center denotes the backdoor target.

\begin{figure}[h]
    \centering
    \includegraphics[width=0.55\linewidth]{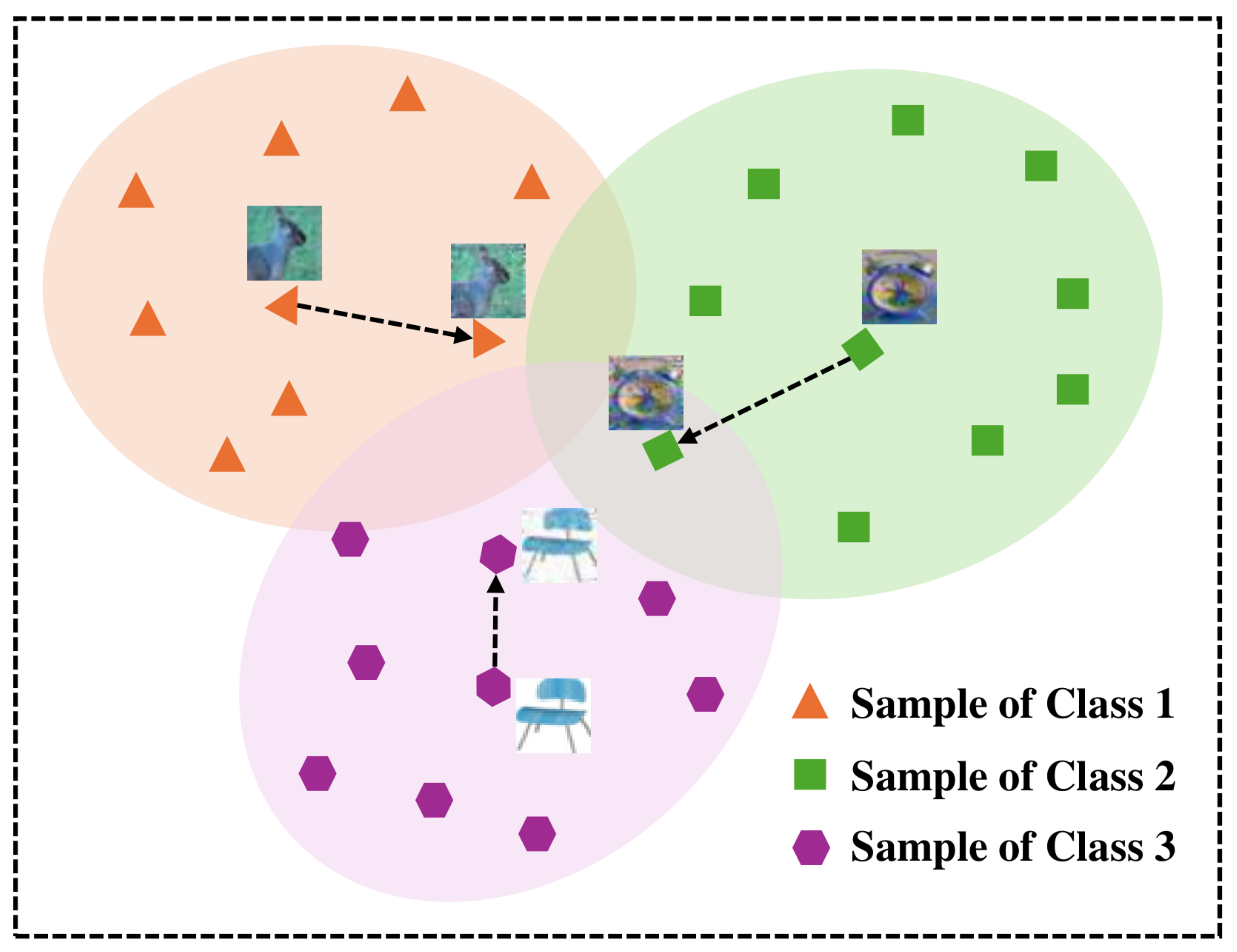}
    \caption{Illustration of feature space alignment using \textit{ShadowPrint}. The diagram showcases how poisoned samples are clustered in the feature space, aligning the features of the triggered samples.}
    \label{fig:method}
\end{figure}

However, conventional backdoor attacks complete the process of backdoor learning with model training, thus requiring many poisoned samples. To mitigate this limitation, we exploit the analysis above and propose a clustering-based trigger optimization strategy. Generally speaking, we can find a universal trigger that can help the cluster of poisoned samples before model training, reducing the burden of backdoor learning, as illustrated in Fig.~\ref{fig:method}. The trigger optimization process focuses on manipulating the feature embeddings in the last fully connected (FC) layer, to align the feature embeddings of poisoned samples effectively in the feature space. 

Specifically, the optimization process relies on the following custom loss to maximize alignment between feature vectors of poisoned samples in the model’s embedding space:




\begin{equation}
\label{eq:loss}
\mathcal{L}_\text{cluster} = \frac{1}{N^2} \sum_{i,j \, :\, i \neq j} \frac{Z_i \cdot Z_j^T}{\|Z_i\| \|Z_j\|}
\end{equation}
where, $Z_i$ and $Z_j$ represent the feature vectors of poisoned samples $x'_i$ and $x'_j$ via $f_\text{adv}^{fc}$, i.e., the model’s last FC layer, $N$ denoting the size of samples. By minimizing~\eqref{eq:loss}, the optimization process encourages higher cosine similarity between these feature vectors, effectively clustering them closer together in the embedding space. Consequently, as outlined in Algorithm~\ref{alg:ShadowPrint}, the optimized trigger $t$ iteratively aligns the features $Z$ of all poisoned samples to a common cluster center, enhancing the stealthiness and robustness of the backdoor attack. This clustering-based trigger optimization strategy reduces the need for a large number of poisoned samples, thereby improving the efficiency of the attack.

\begin{algorithm}
\caption{Optimization Process for ShadowPrint}
\label{alg:ShadowPrint}
\begin{algorithmic}[1]
\State \textbf{Input:} Attacker's Dataset $\mathcal{D}_\text{adv}$; Attacker's Model $f_\text{adv} $; Optimizing Steps $K $
\State  $t \gets \mathcal{N}(0,0.5)$
\For{epoch in $K $}
    \For{each batch $(x, y) \in \mathcal{D}_\text{adv} $}
        \State $x' \gets T(x, t) $\Comment{Obtain triggered samples.}
        \State $Z \gets f_\text{adv}^{fc}(x')$ \Comment{Capture the embeddings}. 
        \State Update $t$ with $\nabla_{t}\mathcal{L}_\text{cluster}$\Comment{Use Adam optimizer.}
    \EndFor
\EndFor
\State \textbf{Return:} Generated trigger $t $.
\end{algorithmic}
\end{algorithm}

\begin{table*}[htbp]
\centering
\tabcolsep=0.1cm
\renewcommand{\arraystretch}{1}
\caption{Dirty Label Attack. We evaluate \textit{ShadowPrint} under the dirty-label attack mode by testing the CA and ASR across different poison rates and attacker settings. } 
\label{tab:dirty_label}
\resizebox{\textwidth}{!}{%
\begin{threeparttable}
\begin{tabular}{@{}cccccccccccccc@{}}
\toprule
\multirow{2}{*}{\shortstack{\textbf{Target} \\ \textbf{Model}}} & \multirow{2}{*}{\shortstack{\textbf{Poison} \\ \textbf{Ratio}}} & \multicolumn{4}{c}{\textbf{CIFAR-10}} & \multicolumn{4}{c}{\textbf{CIFAR-100}} & \multicolumn{4}{c}{\textbf{TinyImageNet}} \\
\cmidrule(l){3-14} 
 &  & \textbf{\textit{Baseline}} & \textbf{\textit{ResNet18}} & \textbf{\textit{ResNet34}} & \textbf{\textit{VGG13BN}} & \textbf{\textit{Baseline}} & \textbf{\textit{ResNet18}} & \textbf{\textit{ResNet34}} & \textbf{\textit{VGG13BN}} & \textbf{\textit{Baseline}} & \textbf{\textit{ResNet18}} & \textbf{\textit{ResNet34}} & \textbf{\textit{VGG13BN}} \\
\midrule
\multirow{2}{*}{ResNet18} & 0.0001 & \multirow{2}{*}{0.920} & \underline{0.919/0.994} & 0.919/0.999 & 0.919/0.997 & \multirow{2}{*}{0.690} & \underline{0.687/0.993} & 0.688/0.995 & 0.692/0.999 & \multirow{2}{*}{0.508} & \underline{0.456/0.997} & 0.454/0.999 & 0.462/0.998 \\
 & 0.0005 &  & \underline{0.922/1.000} & 0.919/1.000 & 0.921/1.000 &  & \underline{0.692/1.000} & 0.695/0.999 & 0.688/0.998 &  & \underline{0.462/1.000} & 0.458/1.000 & 0.462/1.000 \\
\midrule
\multirow{2}{*}{ResNet34} & 0.0001 & \multirow{2}{*}{0.925} & 0.922/0.999 & \underline{0.922/0.997} & 0.924/1.000 & \multirow{2}{*}{0.702} & 0.701/0.996 & \underline{0.701/0.993} & 0.700/1.000 & \multirow{2}{*}{0.525} & 0.449/0.997 & \underline{0.481/1.000} & 0.480/0.981 \\
 & 0.0005 &  & 0.921/1.000 & \underline{0.925/1.000} & 0.924/1.000 &  & 0.701/1.000 & \underline{0.700/1.000} & 0.701/0.999 &  & 0.462/1.000 & \underline{0.456/1.000} & 0.467/1.000 \\
\midrule
\multirow{2}{*}{VGG13BN} & 0.0001 & \multirow{2}{*}{0.919} & 0.920/0.994 & 0.920/1.000 & \underline{0.918/1.000} & \multirow{2}{*}{0.701} & 0.704/0.998 & 0.706/0.997 & \underline{0.698/0.997} & \multirow{2}{*}{0.493} & 0.459/0.997 & 0.467/0.998 & \underline{0.460/0.998} \\
 & 0.0005 &  & 0.921/1.000 & 0.915/1.000 & \underline{0.920/1.000} &  & 0.699/0.999 & 0.703/0.999 & \underline{0.698/0.999} &  & 0.457/1.000 & 0.462/1.000 & \underline{0.463/1.000} \\
\bottomrule
\end{tabular}%
 \begin{tablenotes}   
    \footnotesize
    \item [1]\textit{Note:} For Table~\ref{tab:dirty_label}, \ref{tab:clean_label}, \ref{tab:train_scale}, and \ref{tab:trigger_weight}: Each cell contains two values: CA / ASR.
    \item [2] The columns represent different attack optimization models (e.g., ResNet18, ResNet34, VGG13BN) for the corresponding dataset.
    \item [3] The underlined cells correspond to the white-box attacker A1, while the remaining cells correspond to the black-box attacker A2.
\end{tablenotes}
\end{threeparttable}
}
\vspace{-1.2em}
\end{table*}

\subsubsection{Stage 2-Attack Execution}



The purpose of \textit{ShadowPrint} is to encompass a broader spectrum of attacker types in real-world application scenarios, as we have previously described in Section \ref{Threat Model}. Consequently, we propose three distinct attack modes under \textit{ShadowPrint}:

\begin{itemize}
    \item \textbf{Dirty Label Attack:} 
    Dirty label attacks manipulate the samples and labels simultaneously. 
    \item \textbf{Clean Label Attack:} Samples from the target class are manipulated while labels remain unchanged, ensuring more practical and stealth-oriented application scenarios. 
    \item \textbf{Data-Free Attack:} Using auxiliary data from other domains to train surrogate model $f_\text{adv}$ and construct $\mathcal{D}_\text{adv}$ (i.e., $f_\text{adv} \neq f$ and $\mathcal{D}_\text{adv} \cap \mathcal{D} = \emptyset$).     
\end{itemize}

Note that, for Scenario A3, the assumption is that the attacker does not have access to the training data or model architecture. This is a specific and challenging scenario where information about the target model is unavailable. To address this, the data-free attack builds upon the principles of dirty-label attacks by leveraging auxiliary datasets and models from other domains.



\begin{figure}[h]
    \centering
    \includegraphics[width=0.45\linewidth]{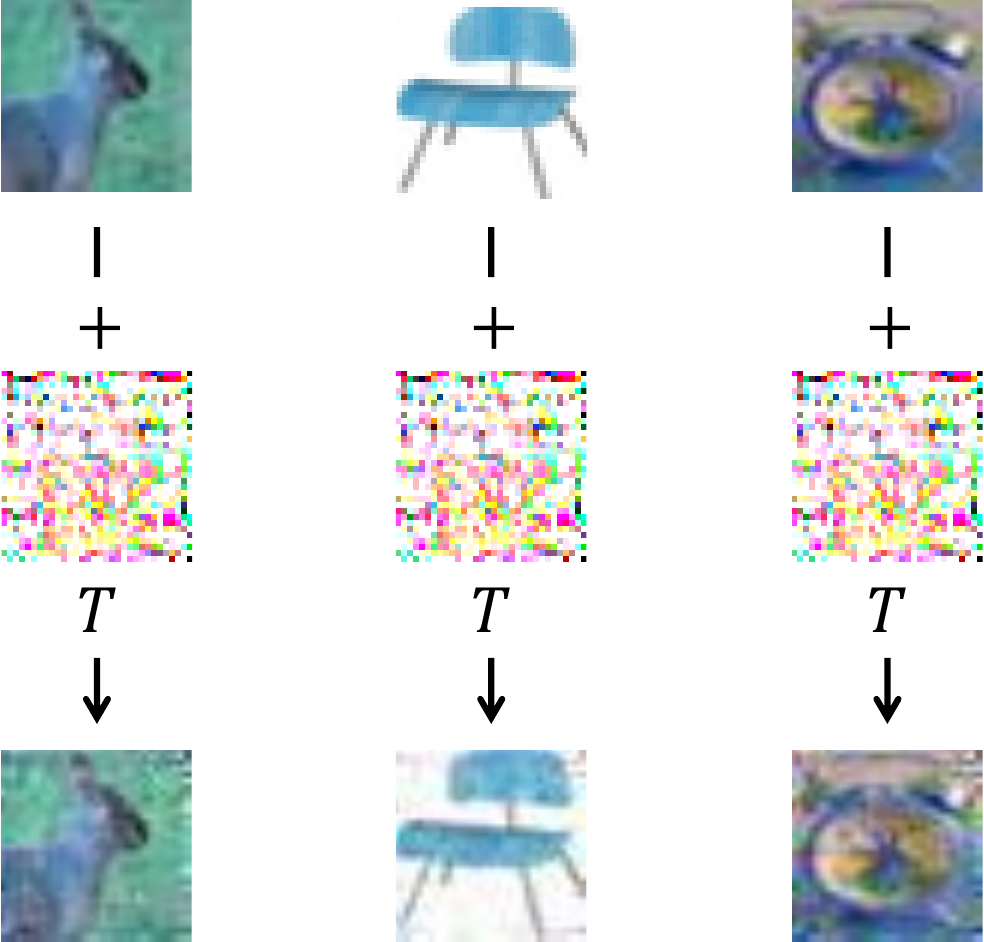}
    \caption{Visualization of the poisoning process. The visualization depicts poisoned samples created using the backdoor trigger.}
    \label{fig:visualization}
\end{figure}

\textit{ShadowPrint} sets itself apart from many SOTA backdoor attacks by employing a remarkably low poison rate while retaining its ability to execute multiple attack types.
As shown in Fig.~\ref{fig:visualization}, 
the trigger induces subtle differences between poisoned and clean samples, ensuring high stealth and effectiveness. This makes \textit{ShadowPrint} harder for existing defenses to detect, offering a versatile and potent backdoor attack method.

\begin{table*}[h]
\centering
\tabcolsep=0.1cm
\renewcommand{\arraystretch}{1}
\caption{Clean Label Attack. We evaluate \textit{ShadowPrint} under the clean-label attack mode by testing the CA and ASR across different poison rates and attacker settings.}
\label{tab:clean_label}
\resizebox{\textwidth}{!}{%
\begin{tabular}{@{}cccccccccccccc@{}}
\toprule
\multirow{2}{*}{\shortstack{\textbf{Target} \\ \textbf{Model}}} & \multirow{2}{*}{\shortstack{\textbf{Poison} \\ \textbf{Ratio}}} & \multicolumn{4}{c}{\textbf{CIFAR-10}} & \multicolumn{4}{c}{\textbf{CIFAR-100}} & \multicolumn{4}{c}{\textbf{TinyImageNet}} \\ \cmidrule(l){3-14} 
 &  & \textbf{\textit{Baseline}} & \textbf{\textit{ResNet18}} & \textbf{\textit{ResNet34}} & \textbf{\textit{VGG13BN}} & \textbf{\textit{Baseline}} & \textbf{\textit{ResNet18}} & \textbf{\textit{ResNet34}} & \textbf{\textit{VGG13BN}} & \textbf{\textit{Baseline}} & \textbf{\textit{ResNet18}} & \textbf{\textit{ResNet34}} & \textbf{\textit{VGG13BN}} \\ \midrule
\multirow{2}{*}{ResNet18} & 0.0001 & \multirow{2}{*}{0.920} &  \underline{0.923/0.998} & 0.922/0.962 & 0.920/0.890 & \multirow{2}{*}{0.690} & \underline{0.689/0.999} & 0.699/0.998 & 0.689/0.964 & \multirow{2}{*}{0.508} & \underline{0.453/0.999} & 0.459/0.999 & 0.462/0.991 \\
 & 0.0005 &  & \underline{0.926/1.000} & 0.923/1.000 & 0.923/0.999 &  & \underline{0.699/0.999} & 0.693/1.000 & 0.684/1.000 &  & \underline{0.457/1.000} & 0.463/1.000 & 0.461/1.000 \\
\midrule
\multirow{2}{*}{ResNet34} & 0.0001 & \multirow{2}{*}{0.925} & 0.924/0.947 & \underline{0.925/0.999} & 0.924/0.999 & \multirow{2}{*}{0.702} & 0.692/0.999 & \underline{0.689/0.998} & 0.695/0.984 & \multirow{2}{*}{0.525} & 0.468/0.989 & \underline{0.477/0.998} & 0.469/0.996 \\
 & 0.0005 &  & 0.923/1.000 & \underline{0.930/1.000} & 0.926/1.000 &  & 0.697/0.998 & \underline{0.696/0.999} & 0.699/0.997 &  & 0.463/1.000 & \underline{0.469/1.000} & 0.470/1.000 \\
\midrule
\multirow{2}{*}{VGG13BN} & 0.0001 & \multirow{2}{*}{0.919} & 0.920/0.943 & 0.919/1.000 & \underline{0.921/0.999} & \multirow{2}{*}{0.701} & 0.703/1.000 & 0.704/0.994 & \underline{0.702/0.986} & \multirow{2}{*}{0.493} & 0.459/0.999 & 0.461/1.000 & \underline{0.465/0.998} \\
 & 0.0005 &  & 0.920/1.000 & 0.918/1.000 & \underline{0.916/1.000} &  & 0.696/0.996 & 0.701/0.999 & \underline{0.702/0.986} &  & 0.462/1.000 & 0.461/1.000 & \underline{0.461/1.000} \\
\bottomrule
\end{tabular}%
}
\vspace{-1.2em}
\end{table*}
\begin{table}[h]
\centering
\tabcolsep=0.1cm
\renewcommand{\arraystretch}{1}
\caption{Data-Free Attack}
\label{tab:data_free}
\begin{tabular}{llll}
\toprule
\textbf{Target Model} & \textbf{Target Dataset} & \textbf{Baseline} & \textbf{CA/ASR} \\ 
\midrule
\multirow{2}{*}{ResNet18} & CIFAR-10 & 0.920 & 0.908/0.913 \\
 & TinyImageNet & 0.508 & 0.458/1.000 \\ 
 \midrule
\multirow{2}{*}{VGG13BN} & CIFAR-10 & 0.919 & 0.903/0.859 \\
 & TinyImageNet & 0.493 & 0.465/1.000 \\ 
\bottomrule
\end{tabular}
\vspace{-1.2em}
\end{table}

\section{Evaluation}

\subsection{Experimental Settings}

\textbf{Datasets and Models.} We evaluate \textit{ShadowPrint} on three commonly used models: ResNet18~\cite{he2016deep}, ResNet34, and VGG13BN~\cite{simonyan2014very}. The datasets used for evaluation include CIFAR-10~\cite{krizhevsky2009learning}, CIFAR-100~\cite{krizhevsky2009learning}, and TinyImageNet~\cite{le2015tiny}, which are widely adopted for image classification tasks. The models are trained on these datasets, and the backdoor attack is introduced by poisoning the training data at varying rates.


\textbf{Attack Settings.} In our evaluation, we experiment with different poisoning rates (e.g., $0.01\%$ and $0.05\%$, much lower than common $0.1\%$ or higher settings), attack methods (e.g., clean-label and dirty-label poisoning, wider variety than other single-method attack), and attacker's capabilities (e.g., Scenario A1, A2, and A3, more types of Scenario Assumption and lower capability configurations than other attacks).
We also study the impact of various factors, such as trigger weight and the scale of the attacker's dataset (i.e., train scale), on the attack's success. The poison rate is defined as the fraction of poisoned samples in the dataset, while the trigger weight refers to the strength of the trigger’s influence in the backdoor attack, as detailed in \eqref{eq:blended}.


\textbf{Evaluation Metrics.} We measure the following key metrics to evaluate the effectiveness and stealthiness of \textit{ShadowPrint}:
\begin{itemize}
    \item \textbf{Attack Success Rate (ASR):} The percentage of poisoned samples that are misclassified to the target label. Higher ASR indicates better attack effectiveness.
    \item \textbf{Clean Accuracy (CA):} The model accuracy on clean samples under the attack. Higher CA indicates a lower impact on the model's usability by the attack, thereby demonstrating better stealthiness of the attack. 
    \item \textbf{Defense Detection Rate (DDR):} The effectiveness of \textit{ShadowPrint} in evading state-of-the-art defense mechanisms. The value ranges from $0$ to $1$, where a lower value indicates a better evasion of defense mechanisms.
\end{itemize}

\subsection{Attack Effectiveness and Stealthiness}



\textbf{Dirty-Label \& Clean-Label Attack.} We evaluate \textit{ShadowPrint} in both dirty-label and clean-label setting under the capability of Scenario A1 and A2, using CIFAR-10, CIFAR-100, and TinyImageNet as evaluation datasets. In the dirty-label attack setting, where the training dataset contains mislabeled samples, \textit{ShadowPrint} achieves high ASRs across all datasets, as shown in Table~\ref{tab:dirty_label}, while maintaining high CA, demonstrating its stealthiness and minimal disruption to benign input performance. Similarly, in the clean-label attack setting, where all training samples are correctly labeled, \textit{ShadowPrint} achieves comparable ASRs with little to no impact on CA, as evidenced in Table~\ref{tab:clean_label}. Finally even with a low poison rate, \textit{ShadowPrint} effectively disrupts the model’s decision-making process in both settings, showcasing its versatility and effectiveness in evading detection. For instance, with a poison rate of $0.05\%$, \textit{ShadowPrint} attains ASRs exceeding 95\% while retaining CA above 92\%, highlighting its balance between attack effectiveness and stealth.

\textbf{Data-Free Attack.} We also test \textit{ShadowPrint} in a data-free scenario, where the attacker has no access to the training data (i.e., Scenario A3). Under this setting, We assume that the attacker employs a surrogate model ($f_\text{adv} = \text{ResNet34}$) and auxiliary data ($\mathcal{D}_\text{adv} = \text{CIFAR100}$)
for an approximation. Table~\ref{tab:data_free} shows the evaluation of the robustness of \textit{ShadowPrint} under limited access conditions. Despite the lack of access to the training data, \textit{ShadowPrint} still manages to perform a successful attack, with a reasonable ASR and minimal impact on CA. The results highlight the flexibility and robustness of \textit{ShadowPrint} in varying attacker scenarios.

\subsection{Evasion of Backdoor Detection}
\textit{ShadowPrint} is evaluated against SOTA defense mechanisms, including IBD-PSC~\cite{hou2024ibd}, SCALE UP~\cite{guo2023scale}, and Beatrix~\cite{ma2022beatrix}. Table~\ref{tab:defense} shows that \textit{ShadowPrint} achieves low DDRs across all attacker scenarios. For example, under Scenario A1, \textit{ShadowPrint} records DDR values as low as 0.05, outperforming comparable methods and demonstrating its ability to evade detection while maintaining high ASRs and CA. These results underscore the method’s stealth and effectiveness, even against robust defensive measures.

We evaluate the effectiveness of \textit{ShadowPrint} against several SOTA defense mechanisms designed to detect backdoor attacks. These include IBD-PSC~\cite{hou2024ibd}, SCALE UP~\cite{guo2023scale}, and Beatrix~\cite{ma2022beatrix}. As shown in Table~\ref{tab:defense}, under all attacker's scenarios, \textit{ShadowPrint} successfully records DDR values lower than $0.12\%$, outperforming comparable methods and demonstrating its ability to evade detection while maintaining high ASRs and CA. These results underscore the method’s stealth and effectiveness, even against robust defensive measures.

\begin{table}[h]
\centering
\tabcolsep=0.1cm
\renewcommand{\arraystretch}{1}
\caption{Train Scale Study}
\label{tab:train_scale}
\begin{tabular}{@{}cccccccc@{}}
\toprule
\multirow{2}{*}{\shortstack{\textbf{Target} \\ \textbf{Model}}} & \multirow{2}{*}{\shortstack{\textbf{Train} \\ \textbf{Scale}}} & \multicolumn{4}{c}{\textbf{CIFAR-10}} \\
\cmidrule(l){3-6} 
 &  & \textbf{\textit{Baseline}} & \textbf{\textit{ResNet18}} & \textbf{\textit{ResNet34}} & \textbf{\textit{VGG13BN}} \\
\midrule
\multirow{4}{*}{ResNet18} & 0.1 & \multirow{4}{*}{0.920} & \underline{0.923/0.933} & 0.922/0.992 & 0.925/0.992 \\
 & 0.2 &  & \underline{0.924/0.999} & 0.923/0.999 & 0.921/1.000 \\
 & 0.3 &  & \underline{0.926/0.999} & 0.923/1.000 & 0.923/1.000 \\
 & 0.5 &  & \underline{0.925/0.999} & 0.921/1.000 & 0.921/1.000 \\
\midrule
\multirow{4}{*}{ResNet34} & 0.1 & \multirow{4}{*}{0.925} & 0.921/0.973 & \underline{0.927/0.966} & 0.922/0.988 \\
 & 0.2 &  & 0.925/1.000 & \underline{0.924/0.998} & 0.925/1.000 \\
 & 0.3 &  & 0.926/1.000 & \underline{0.920/0.999} & 0.922/1.000 \\
 & 0.5 &  & 0.926/1.000 & \underline{0.920/1.000} & 0.922/1.000 \\
\midrule
\multirow{4}{*}{VGG13BN} & 0.1 & \multirow{4}{*}{0.919} & 0.919/0.981 & 0.917/0.975 & \underline{0.917/0.998} \\
 & 0.2 &  & 0.923/1.000 & 0.917/0.997 & \underline{0.918/1.000} \\
 & 0.3 &  & 0.920/1.000 & 0.916/1.000 & \underline{0.920/1.000} \\
 & 0.5 &  & 0.918/1.000 & 0.920/1.000 & \underline{0.917/1.000} \\
\bottomrule
\end{tabular}
\vspace{-1.2em}
\end{table}

\subsection{Ablation Study}
To further understand the behavior of \textit{ShadowPrint}, we conduct an ablation study by evaluating the attack’s performance under various configurations. Specifically, we investigate the effect of the poison rate, trigger weight, and the scale of the attacker's dataset on the attack’s success and stealth.

\textbf{Poison Rate.} Table~\ref{tab:clean_label} and Table~\ref{tab:dirty_label} show the impact of different poison rates on ASR and CA in different settings. As expected, increasing the poison rate results in a higher ASR. However, \textit{ShadowPrint} maintains its high stealth even at higher poison rates, as evidenced by the minimal impact on CA. This stability highlights the stealthiness of the method and its ability to minimize disruption to clean inputs. 

\textbf{Trigger Weight.} We also study the impact of various trigger weights.
As shown in Table~\ref{tab:trigger_weight}, the ASR increases as the trigger weight is adjusted, but the CA remains stable. The ablation study demonstrates that the attack can be finely tuned to balance attack success and stealth.

\textbf{Train Scale.} In Table~\ref{tab:train_scale}, we evaluate the effect of varying the scale of the attacker’s dataset. As expected, the ASR increases with larger training scales, but the model’s accuracy on clean samples remains unaffected. This study shows that \textit{ShadowPrint} performs robustly across different levels of attacker knowledge and dataset sizes.

\begin{table}[h]
\centering
\tabcolsep=0.1cm
\renewcommand{\arraystretch}{1}
\caption{Trigger Weight Study}
\label{tab:trigger_weight}
\begin{tabular}{@{}cccccccc@{}}
\toprule
\multirow{2}{*}{\shortstack{\textbf{Target} \\ \textbf{Model}}} & \multirow{2}{*}{\shortstack{\textbf{Trigger} \\ \textbf{Weight}}} & \multicolumn{4}{c}{\textbf{CIFAR-10}} \\
\cmidrule(l){3-6} 
 &  & \textbf{\textit{Baseline}} & \textbf{\textit{ResNet18}} & \textbf{\textit{ResNet34}} & \textbf{\textit{VGG13BN}} \\
\midrule
\multirow{4}{*}{ResNet18} & 0.1 & \multirow{4}{*}{0.920} & \underline{0.920/0.395} & 0.924/0.415 & 0.921/0.294 \\
 & 0.2 &  & \underline{0.925/0.998} & 0.919/0.994 & 0.922/1.000 \\
 & 0.3 &  & \underline{0.923/1.000} & 0.923/1.000 & 0.924/1.000 \\
 & 0.5 &  & \underline{0.921/1.000} & 0.923/1.000 & 0.924/1.000 \\
\midrule
\multirow{4}{*}{ResNet34} & 0.1 &\multirow{4}{*}{0.925} & 0.924/0.464 & \underline{0.922/0.500} & 0.924/0.427 \\
 & 0.2 &  & 0.925/0.991 & \underline{0.925/0.999} & 0.924/1.000 \\
 & 0.3 &  & 0.927/0.999 & \underline{0.929/0.999} & 0.924/1.000 \\
 & 0.5 &  & 0.923/1.000 & \underline{0.925/1.000} & 0.925/1.000 \\
\midrule
\multirow{4}{*}{VGG13BN} & 0.1 & \multirow{4}{*}{0.919} & 0.921/0.200 & 0.920/0.568 & \underline{0.921/0.211} \\
 & 0.2 &  & 0.917/1.000 & 0.920/0.996 & \underline{0.919/1.000} \\
 & 0.3 &  & 0.918/1.000 & 0.920/1.000 & \underline{0.923/1.000} \\
 & 0.5 &  & 0.923/1.000 & 0.917/1.000 & \underline{0.917/1.000} \\
\bottomrule
\end{tabular}
\vspace{-1.2em}
\end{table}
\subsection{Discussion}

\textit{ShadowPrint} proves to be an effective and stealthy backdoor attack, achieving high ASR while maintaining strong CA. Unlike traditional attacks that manipulate input visuals, \textit{ShadowPrint} targets internal feature representations, making it more resistant to detection and defenses. The ablation study shows that it is highly adaptable, balancing effectiveness and stealth through hyperparameters. 
Even in data-free scenarios, \textit{ShadowPrint} performs well, indicating its potential for real-world applications. However, its reliance on feature space manipulation suggests the need for novel and effective detection strategies to counter such attacks.

\begin{table}[h]
\centering
\tabcolsep=0.1cm
\renewcommand{\arraystretch}{1}
\caption{Defense Study}
\label{tab:defense}
\begin{tabular}{@{}ccccc@{}}
\toprule
\textbf{Label} & \textbf{Scenario} & \textbf{IBD\_PSC} & \textbf{SCALE\_UP} & \textbf{Beatrix} \\ \midrule
\multirow{3}{*}{DIRTY} & White-Box & 0.009 & 0.000 & 0.054 \\
 & Black-Box & 0.004 & 0.000 & 0.057 \\
 & Data-Free & 0.004 & 0.000 & 0.057 \\ \midrule
\multirow{3}{*}{CLEAN} & White-Box & 0.010 & 0.000 & 0.061 \\
 & Black-Box & 0.046 & 0.000 & 0.052 \\
 & Data-Free & 0.117 & 0.000 & 0.063 \\ \bottomrule
\end{tabular}
\vspace{-1.2em}
\end{table}

\section{Conclusion}
In this paper, we presented \textit{ShadowPrint}, a novel backdoor attack that manipulates feature embeddings within a model’s feature space to achieve both high attack success and stealth. 
Leveraging existing method limitations, \textit{ShadowPrint} reduces reliance on strong attacker capabilities and performs well across diverse scenarios.
Experimental results also demonstrate  \textit{ShadowPrint} 
excels by effectively disrupting model performance while maintaining high accuracy on clean samples.

\section*{Acknowledgment}
This work was partly supported by the NSFC under No. U244120033, U24A20336, 62172243, 62402425 and 62402418, the China Postdoctoral Science Foundation under No. 2024M762829, the Zhejiang Provincial Natural Science Foundation under No. LD24F020002, and the Zhejiang Provincial Priority- Funded Postdoctoral Research Project under No. ZJ2024001.


\bibliographystyle{IEEEtran}

\end{document}